\documentstyle[preprint,aps]{revtex}
\begin{document}
\draft
\title{Horizon problem in a closed universe\protect\\
dominated by fluid with negative pressure}
\author{Jerzy Stelmach \footnote{E-mail: Jerzy\_Stelmach@univ.szczecin.pl}}
\address{Institute of Physics, University of Szczecin, 
 Wielkopolska 15, 70-451 Szczecin, Poland}
\date{\today}
\maketitle
\begin{abstract}
We discuss the horizon problem in a universe dominated by fluid with
negative pressure. We show that for generally accepted value of
nonrelativistic matter energy density parameter $\Omega_{m0}<1$, the
horizon problem can be solved only if the fluid influencing negative
pressure (the so-called ``X'' component) violates the point-wise strong
energy condition and if its energy density is sufficiently large
$(\Omega_{X 0}>1)$. The calculated value of the $\Omega_{X0}$ parameter
allowing for the solution of the horizon problem is confronted with some
recent observational data. Assuming that $p_X/\rho_X<-0.6$ we find that
the required amount of the ``X'' component is not ruled out by the
supernova limits. Since the value of energy density parameter $\Omega_{v0}$
for cosmological constant larger than 1 is excluded by gravitational lensing
observations the value of the ratio $p_X/\rho_X$ should lie between the
values $-1$ and $-0.6$ if the model has to be free of the horizon problem
beeing at the same time consistent with observations. The value of
$\Omega_{X0}+\Omega_{m0}$ in the model is consistent with the constraints
$0.2<\Omega_{\text{tot}}<1.5$ following from cosmic microwave background
observations provided that $\Omega_{m0}$ is low ($<0.2$). 
\end{abstract}

\section{Horizon problem in standard cosmology}

According to the standard scenario, about 300 000 years after the Big
Bang the Universe cooled down to the level that atoms could form.
Electrons were captured by nuclei and photons, which until that time
were in thermal equilibrium with plasma, lost charged partners for
interaction and started their free travel across the Universe. These
photons are detected today and form the so-called cosmic microwave
background (CMB). The radiation has a black-body spectrum and corresponds
to the temperature  $T=2.73\pm 0.01$ K, independently on the
direction it comes from. Extreme isotropy of the radiation is its most
puzzling property because the regions from which two hitting us today
antipodal photons come, could never communicate with each other. Hence
the same temperature of these regions cannot be explained in the
standard scenario. This is called a horizon problem. Its
inevitability in the standard cosmological model is usually illustrated
by non-intersection of two past light cones of these regions at the
recombination epoch, i.e. at the epoch when the relic radiation appeared
(Fig.\ \ref{fig1}). This non-intersection means that there were no events 
in the history of the universe which could influe the recombination process 
at the points $A$ and $B$ simultaneously.

\section{Some attempts to solve the horizon problem}

One of the first conclusions which follow even from superficial analysis
of the picture (Fig.\ \ref{fig1}) is that the horizon problem exists 
because the recombination of atoms took place so early comparing to the 
age of the universe ($t_R/t_0<10^{-4}$). If $t_R$ was large enough the 
appropriate light cones would intersect and the problem would not appear.
Unfortunately in the framework of the standard scenario the recombination
epoch cannot be shifted significantly into the future. What we can do
instead is to assume that the last scattering of relic photons did not
take place at $t_R$ but more recently, for example, due to the Compton
scattering on free electrons filling up the whole universe \cite{wein72}.
Detailed calculations show, however, that for generally accepted values of
cosmological parameters the horizon problem still exists \cite{stel90}.
In other words shifting the last scattering surface into the future
does not help to solve the problem.

Careful glance at the picture suggests another solution to the
problem: an appropriate bending of past light cones of the points $A$
and $B$  (concave form instead of convex one) can cause that the
horizons would overlap (Fig.\ \ref{fig2}). This is realized in an 
inflationary scenario \cite{guth81} where a large cosmological term gives 
the required form of the light cones. Inspite of the fact that there is no
unique physical model of the inflation the idea of rapid exponential or
power law expansion of the early universe is regarded as very attractive
because it solves additionally several other problems of the standard
model. We shall not discuss this topic now. In inflationary scenarios the
horizon problem does not appear because the past light cones of points $A$
and $B$ bend radically towards the time axis (Fig.\ \ref{fig2}) due to the
accelerating expansion.

In the present paper we discuss the possibility
of solving the horizon problem without very effective inflationary epoch
in the early universe and without shifting the last scattering surface
into the future. The possibility arises in the closed universe, in which
relic photons may come from geometrical antipode. This is of course
trivial fact, but straightforward calculation shows that without fluid
with negative pressure such universe should be extremely dense, what
cannot be accepted. Inspite of the fact that many observations favor
open or flat universe, the closed one is still not ruled out
\cite{white96}. Since the universe in our model, apart from the
relativistic and nonrelativistic matter, is also filled with the fluid
influencing negative pressure (simulating repulsive force) the evolution
of resulting closed cosmological models do not end up with the Big Crunch.
Contrary, the universe is ever expanding reaching the size consistent
with observations. Solution of the horizon problem in such models was
discussed by several authors. Especially two kinds of fluid, in the
context of the horizon problem, were in the past taken into account: the
so-called string-like fluid, described by the equation of state
$p_s=-\rho_s/3$ \cite{dav87,stel94,kam96}, and the cosmological constant
\cite{white96}. It seems that the first kind of fluid (string-like one)
is already ruled out by supernova observations \cite{garn98} which
indicate that the ratio $\alpha_X$ of the pressure $p_X$ to the energy
density $\rho_X$ of the unknown, the so-called ``X'' component of the
universe must be less than $-0.6$ (95\% confidence). As regards the
second kind of fluid, most of recent astronomical observations suggest
existence of positive cosmological constant
\cite{fort98,filip98,cooray98}, but the upper limit for its density is slightly too low in
order to solve the horizon problem, especially if the nonrelativistic
matter content of the universe is large \cite{baum98}. 

The purpose of the present paper is discussion of the horizon
problem in a universe dominated by a more general kind of fluid
influencing negative pressure.

In the next section we shortly motivate dealing with the fluid with
negative pressure.

In Section IV we discuss the horizon problem in the FRW model with 
such fluid.

In Section V we show that the horizon problem does not appear if the
fluid violates the point-wise strong energy condition and if its
contribution to the total energy density of the universe is predominant,
closing the universe. Two special cases: vacuum and string-like matter
dominated models, are discussed in more detail. In the last section we
compare the models with observations and summarize the results.

\section{Motivation for dealing with fluid with negative pressure}

Introduction of fluid with negative pressure in cosmology has a long
history. Einstein introduced it in form of the $\Lambda$-term in order
to get static cosmological model by compensating gravitational atraction
with repulsive affect of the cosmological constant. In recent years
cosmological models with $\Lambda$-term have been intensively
investigated in the context of large scale structure formation
\cite{klyp97} as well as in the context of the age-of-the-universe
problem \cite{kraus97,sper97}. In both cases vacuum energy forms a
smooth component of the dark matter. Visser considered the
age-of-the-universe problem by treating it, as far as possibly, in a
model independent way, without assuming any particular equation of state,
and he showed that if the Hubble parameter is high enough, in order to
solve the problem, the point-wise strong energy condition (SEC) must be
violated between the epoch of galaxy formation and the present
\cite{vis97}. Another reason for dealing with the ``X'' component
follows from high redshift supernovae observations which indicate that
the universe is accelerating \cite{kim98,riess98}. The simplest way to
explain this is to assume that the universe is dominated by some fluid
with negative pressure. Moreover the observed number of gravitational
lensing events cannot be explained without existence of such form of
matter \cite{cooray98}.

Physical interpretation can be attached to the fluid with negative
pressure according to different physical models. The cosmological
constant interpreted as an energy density of physical vacuum (satisfying
the equation of state $p_v=-\rho_v$) is the most obvious example. The
equation of state $p_s=-\rho_s/3$  corresponds to the string-like matter,
which can be regarded as a network of cosmic strings conformally stretched
by the expansion \cite{vil84,kam96}, global texture \cite{dav87} or
decaying cosmological constant \cite{silv94}. Intermediate equations of
state different from $p_v=-\rho_v$ and $p_s=-\rho_s/3$ can also be
achieved  in models with fundamental fields (scalar, vector, or tensor)
forming on average some ``not normal'' fluid \cite{cald98}.

In the present paper we consider fluid with negative pressure by
assuming the equation of state
\begin{equation}
p_X=\alpha_X\rho_X,
\end{equation}
where
\begin{equation}
-1\leq\alpha_X<-{1\over 3},
\end{equation}
hence violating SEC. Cosmological constant corresponds to $\alpha_v=-1$
and is, in some sense, an extreme possibility. For the purposes of the
present paper we admit, however, the upper bound of the interval
($\alpha_s=-1/3$) as well. Summarizing, we assume that the pressure of the
fluid fulfills the inequality
\begin{equation}
-\rho_X\leq p_X\leq -{1\over 3}\rho_X.
\end{equation}
Following Visser \cite{vis97} we shall call the fluid satisfying the
inequality (3) -- ``abnormal''.

\section{Horizon problem in the FRW model with ``abnormal'' fluid}

We assume that the universe is filled with relativistic matter (e.g.
relic radiation), nonrelativistic matter (e.g. galaxies) and the
``abnormal'' fluid (e.g. cosmological constant, stringlike matter,
etc.) -- the fluid satisfying the inequality (3). Let us define (cf.
Fig.\ \ref{fig1}):

$\displaystyle r_0\equiv c\int_0^{t_0}{dt\over R(t)}$ -- comoving radius
of the observer's particle horizon,

$\displaystyle r_R\equiv c\int_0^{t_R}{dt\over R(t)}$ -- comoving
radius of the particle horizon at the recombination epoch,

$\displaystyle \chi_R\equiv r_0-r_R=c\int_{t_R}^{t_0}{dt\over
R(t)}$ -- comoving coordinate of the regions from which \hfill

\hskip 5cm hitting us today relic photons were emitted,

$R(t)$ -- scale factor.

It follows that the horizon problem does not appear if
\begin{equation}
2r_R>r_0, ~~\text{or ~equivalently ~if} ~~ r_R>\chi_R.
\end{equation}
In other words comoving radius of the particle horizon  at the
recombination epoch must be larger than the present comoving distance to
the scattering surface taking place at that epoch.

In the model under consideration the expressions for $r_R$ and
$\chi_R$ may be rewritten in the form (c.f. \cite{stel90})
\begin{eqnarray}
r_R=\left({\Omega_{r0}+\Omega_{m0}+\Omega_{X0}-1\over k}\right)^{1/2}
\int_{z_R+1}^\infty\biggl[x^4\left(\Omega_{r0}+
{\Omega_{m0}\over z_R+1}\right) &+& (1-\Omega_{r0}-\Omega_{m0}-
\Omega_{X0})x^2 \nonumber \\
&+& \Omega_{X0}x^{3(\alpha_X+1)}\biggr]^{-1/2}dx,
\end{eqnarray}
\begin{eqnarray}
\chi_R=\left({\Omega_{r0}+\Omega_{m0}+\Omega_{X0}-1\over k}\right)^{1/2}
\int_{1}^{z_R+1}\Bigl[\Omega_{r0}x^4 + \Omega_{m0}x^3 &+&(1-\Omega_{r0}-
\Omega_{m0}-\Omega_{X0})x^2 \nonumber \\
&+&\Omega_{X0}x^{3(\alpha_X+1)}\Bigr]^{-1/2}dx,
\end{eqnarray}
where $z_R$ -- redshift corresponding to the recombination epoch,

\hskip-6mm$\Omega_{r0}, \Omega_{m0}$ -- relativistic and
nonrelativistic matter energy density parameters,

$\Omega_{X0}$ -- ``abnormal'' fluid energy density parameter.

Since the integrands in both cases are rapidly decreasing functions of $x$,
and since $z_R$ is relatively large ($\approx 1200$) even rough estimate of
numerical values of the above integrals leads to the conclusion that $r_R$
cannot be larger than $\chi_R$ for any reasonable values of $\Omega_{r0}$,
$\Omega_{m0}$ and $\Omega_{X0}$. Numerical integration confirms
this estimate. E.g. for $\Omega_{r0}=0.00004,~\Omega_{m0}=0.3$ and
$\Omega_{v0}\approx 0.7$ ($\alpha_X=-1$ -- nearly flat model with
cosmological constant) we get the value $r_R/\chi_R\approx 0.0150$, much
too low to solve the problem. For $\Omega_{m0}=0.1$ and $\Omega_{v0}=0.9$
we get $r_R/\chi_R\approx 0.0153$, and the problem still remains.
Considering other forms of ``abnormal'' fluid ($\alpha_X > -1$) does not
change the ratio significantly. Hence the conclusion could be drawn that
the horizon problem cannot be solved in the standard (noninflationary)
cosmological scenario independently of whether the ``abnormal'' fluid is
involved or not. In the next part of the paper I will show that this
conclusion is not quite correct.

\section{Horizon problem in a closed ``abnormal'' fluid-dominated universe}

Almost constant and small value of the ratio $r_R/\chi_R\approx 0.015$
suggests that the horizon problem cannot be solved in the standard
scenario even with some form of ``abnormal'' fluid. Very small value
of this ratio means that the angle $\theta$ at which we observe today
causally connected region at the recombination epoch -- the so-called
particle horizon \cite{rind56} is of order of only few degrees
($\theta\leq 3^\circ)$ \cite{wein72}. It is known, however, that the
above statement must be revised in a closed cosmological model.

Now let us assume that the universe is closed ($k=1$), i.e. the
total energy density of matter filling up the universe, including
radiation, nonrelativistic matter and the ``abnormal'' fluid is larger
than the critical density. In this case the expressions for $r_R$ and
$\chi_R$ are
\begin{eqnarray}
r_R=(\Omega_{r0}+\Omega_{m0}+\Omega_{X0}-1)^{1/2}\int_{z_R+1}
^\infty\biggl[x^4\left(\Omega_{r0}+{\Omega_{m0}\over z_R+1}\right)
&-&(\Omega_{r0}+\Omega_{m0}+\Omega_{X0}-1)x^2 \nonumber \\
&+&\Omega_{X0}x^{3(\alpha_X+1)}
\biggr]^{-1/2}dx,
\end{eqnarray}
\begin{eqnarray}
\chi_R=(\Omega_{r0}+\Omega_{m0}+\Omega_{X0}-1)^{1/2}\int_1^{z_R+1}
\Bigl[\Omega_{r0}x^4+\Omega_{m0}x^3&-&(\Omega_{r0}+\Omega_{m0}+\Omega_{X0}
-1)x^2 \nonumber \\
&+&\Omega_{X0}x^{3(\alpha_X+1)}\Bigr]^{-1/2}dx,
\end{eqnarray}
and play role of angular coordinates (cf. Fig.\ \ref{fig3}).

As we know from the earlier discussion and also see from the picture
the horizon problem appears (shaded regions do not overlap) if
$\chi_R>r_R$ what is always the case for reasonable values of the
parameters $\Omega_{r0}, \Omega_{m0}$ and $\Omega_{X0}$. However,
closdeness of the universe gives chance to solve the horizon problem even
if $\chi_R\gg r_R$. Note that if $\chi_R$ was very close to $\pi$ then
even for small value of $r_R$ the shaded regions could overlap. Hence
what we need is to fulfill the condition (Fig.\ \ref{fig4})
\begin{equation}
\vert\pi-\chi_R\vert<r_R.
\end{equation}
Since $\Omega_{r0}$ and $\Omega_{m0}$ are more or less fixed (by
observations) we can only vary the ``abnormal'' fluid energy density
parameter $\Omega_{X0}$. Even without performing explicit integrations
we notice that only for
\begin{equation}
-1\leq\alpha_X\leq -{1\over 3}
\end{equation}
$r_R$ as well as $\chi_R$ are growing functions of $\Omega_{X0}$, what
is necessary to approach $\pi$ by $\chi_R$ in order to fulfill the triangle
inequality (10) since $r_R$ is always relatively small. Numerical
calculations show that the growth is relatively fast. Moreover, the fastest
growth is achieved for cosmological constant ($\alpha_v=-1$) which
corresponds to the lower bound of the interval (10), and the slowest one
for the string-like matter ($\alpha_s=-1/3$) -- upper bound of the
interval. Since the cosmological constant acts more effectively let us
focus on this case.

Starting from $\chi_R\approx 0.02$ and $\chi_R\approx 0.0003$ for
$\Omega_{m0}=0.3$ and $\Omega_{v0}=0.7$ we reach $\chi_R\approx 3.11$ and
$r_R\approx 0.04$ for $\Omega_{v0}=1.311$. Note that in the latter case the
condition (9) is fulfilled, and the horizon problem is solved. Further
increasing $\Omega_{v0}$ causes violation of the condition (9) starting
from $\Omega_{v0}=1.327$. In consequence horizon problem appears again.
Summarizing, horizon problem does not appear in the vacuum-dominated closed
universe for $\Omega_{r0}=0.00004,~\Omega_{m0}=0.3$, and $\Omega_{v0}$ from
rather narrow interval
\begin{equation}
1.310<\Omega_{v0}<1.327.
\end{equation}
It is worth mentioning that the condition (10), up to the equality sign
($\alpha_s=-1/3$), is just the violation of the strong energy condition.
We remind that according to Visser \cite{vis97} the same condition had
to be violated in order to solve the age-of-the-universe problem.

As it might have been expected the age of the universe in this case is
relatively large
\begin{equation}
t_0=12.4\times h^{-1}\times 10^9 ~ \text{years,}
\end{equation}
where $h\in (1/2,1)$ is a normalized Hubble constant.  For other values
of $\Omega_{m0}$ (0.015, 0.1 and 0.2) corresponding minimum and maximum
values of $\Omega_{v0}$, for which the horizon problem is solved, are
given in Table \ref{table1}.
The value $\Omega_{m0}=0.015$ in the table is justified by the paper
of Hoell {\it et al.} \cite{hoell94} in which it is argued that
observations of absorption lines of the Lyman $\alpha$ forests of quasars
would suggest that $\Omega_{m0}\approx 0.014$ and $\Omega_{v0}\approx
1.08$. Note that the value $\Omega_{v0}\approx 1.06$ solving the horizon
problem in our model is close to the value of Hoell {\it et al.}
Since the energy density parameters $\Omega_{r}, \Omega_{m}$ and
$\Omega_v$ are not constant in time and the horizon problem is solved
only for the value of $\Omega_{v0}$ belonging to some special interval,
it follows that even if we now lived in the ``isotropic era'' (e.g.
$\Omega_{m0}=0.3, \Omega_{v0}\approx 1.32$) it would not mean that the era
would last forever. In order to see how the horizon problem evolves in
time we evaluate the coordinate $\chi_R(z)$ for a hypothetical observer
living not at the present epoch, but at the epoch determined by the
redshift $z$. The coordinate reads
\begin{equation}
\chi_R(z)=(\Omega_{r0}+\Omega_{m0}+\Omega_{v0}-1)^{1/2}
\int_{z+1}^{z_R+1}\bigl[\Omega_{r0}x^4+\Omega_{m0}x^3
-(\Omega_{r0}+\Omega_{m0}+\Omega_{v0}-1)x^2+\Omega_{v0}\bigr]^{-1/2}dx,
\end{equation}
If we calculate this integral for different values of $z$ starting with
$z_R\approx 1200$ we realize that the horizon problem did not appear until
$z\approx 540$ (for $\Omega_{m0}=0.3$). Nonexistence of the horizon problem
just after the recombination epoch is not surprising of course, because the
observer detects relic photons coming from his closest neighbourhood. The
problem arises some time after the recombination when the observer starts
to detect photons coming from more distant regions, the regions that could
not ever be in thermal equilibrium. And this happens for $z\approx 540$,
i.e. relatively soon after the recombination. In our model the
recombination ($z_R\approx 1200$) takes place about $250\times 10^3\times
h^{-1}$ years after the Big Bang and $z\approx 540$ corresponds to
$880\times 10^3 \times h^{-1}$ years. In the long interval of time between
$z\approx 540$ and $z\approx 0.05$ the horizon problem exists. It
dissapears again about $500\times h^{-1}$ mln of years before the present
epoch and will last for the next $500\times h^{-1}$ mln of years ($z\approx
-0.04$). Before entering the ``isotropic era'' we would observe vanishing of
fluctuations of the microwave background radiation first at the smallest
angular scale and then due to the expansion fluctuations at larger
scales would die out. While leaving the ``isotropic era'' first
fluctuations at largest angular scales ($\theta\approx 180^\circ$) would
come into existence.

As we mentioned before effectiveness of the string-like fluid is
lower than effectiveness of the $\Lambda$-term and larger value of
$\Omega_{s0}$ is required in order to solve the horizon problem. In
Table \ref{table2} minimum and maximum values of $\Omega_{s0}$ (for which
the horizon problem is solved), for different contents of nonrelativistic
matter are presented \cite{stel94}. Note that the age of the universe
in the string-like fluid dominated model is remarkably lower than in the
vacuum dominated case. This might lead to the age-of-the-universe
problem if it turned out that the Hubble constant is large.
\par The discussion presented so far concerned two extreme cases of the
fluid with negative pressure solving the horizon problem in a closed
cosmological model. These cases bound the interval of values of the
$\alpha_X$ -- parameter allowing for the solution of the horizon
problem, from below ($\alpha_v=-1$ -- cosmological constant) and from
above ($\alpha_s=-1/3$ -- string-like matter).

In Fig.\ \ref{fig5} we present admissible values of ``abnormal'' fluid
energy density parameter $\Omega_{X0}$ solving the horizon problem for
various values of the $\alpha_X$ parameter. There are four pairs of lines
corresponding to four values  of the $\Omega_{m0}$ parameter. In each
pair the lower line corresponds to the minimum value of $\Omega_{X0}$
parameter solving the horizon problem, while the upper line -- to the
maximum one. The left bound of the diagram ($\alpha_v=-1$) corresponds to
the cosmological constant and the right bound ($\alpha_s=-1/3$) -- to the
string-like matter. Note that the interval of admissible values of
$\Omega_{X0}$ grows with $\alpha_X$, i.e. it is narrow for the
cosmological constant and relatively large for the string-like matter.
 
\section{Summary, constraints from observations and conclusions}

One of the possibilities of solving the horizon problem in the
standard cosmological scenario (without inflation in the early universe)
arises when we assume that the universe is closed. In such a model two
relic photons reaching us from opposite directions could be in thermal
equilibrium at the recombination epoch if since that time they travelled
almost one-half of the circumference of the universe. Explicit calculation
shows that in a matter-dominated model (without any form of ``not normal''
matter) with a last scattering surface of relic photons taking place at the
recombination epoch ($z_R\approx 1200$) this is possible only in an
extremaly dense universe, which of course cannot be accepted. However, the
situation drastically changes if we admit existence of some form of matter
influencing negative pressure (e.g. cosmological constant, textures,
strings, etc.). Using some estimations it was shown by Davies
\cite{dav87} that in a closed universe with global texture
the horizon problem in fact does not exist. It was also shown
\cite{stel94} that not only textures but any form of the so-called
string-like matter (satisfying the equation of state $p_s=-\rho_s/3$)
solves the horizon problem. In the present paper we showed that any
fluid violating the point-wise strong energy condition, hence described
by the equation of state satisfying the inequality
$-\rho_X\leq p_X\leq -\rho_X/3$, is relevant for solving the horizon problem
as well. Peculiar attention was paid to the lower bound of the interval
(cosmological constant). For four values of the matter energy density
parameter $\Omega_{m0}$ (0.015, 0.1, 0.2 and 0.3) and for the ratio
$p_X/\rho_X$ from the interval $\langle -1, -1/3\rangle$ we calculated
numerical values of the fluid energy density parameter $\Omega_{X0}$
necessary to solve the problem.

The solution of the horizon problem in the model is not forever. It
depends on the epoch of observations. For example the value $\Omega_{v0}
\approx 1.32$ (for $\Omega_{m0}=0.3$) is necessary for observers living
today. They would observe isotropic microwave background radiation for
about $10^9\times h^{-1}$ years. If $\Omega_{v0}$ was larger the
``isotropic era'' would occur earlier; if smaller it would occur in future.

So far we considered theoretical possibility of solving the horizon
problem in a closed universe dominated by fluid with negative pressure
not relating the values of $\Omega_{X0}$ and $\alpha_X$ to observations.
Now we would like to discuss observational constraints on  $\Omega_{X0}$
and $\alpha_X$. We remind that for our purposes the value of $\alpha_X$
must be less than $-1/3$, and the value of $\Omega_{X0}$ must be larger
than 1. There are at least three various methods of measurements which
in recent years are being applied for estimates of global curvature of
the universe. These concern: high-redshift supernovae, gravitational
lensing events and CMB. Especially the first two give strong evidence
that the Universe is accelerating. The simplest way to explain this is
assuming existence of some form of matter with negative pressure. A
candidate which is being taken by many authors most seriously into
account is cosmological constant. However, for our purposes this
candidate is not the best one because gravitational lensing events in
the Hubble Deep Field impose strong upper limit on $\Omega_{v0}$ which
should be remarkably lower than $1$ \cite{cooray98}. The value of this
limit is also confirmed by quasar statistics \cite{koch96}. Another
candidate examined in this context few years ago was string-like matter
\cite{dav87,stel94}, e.g. network of intercommuting cosmic strings,
globally wound texture or decaying $\Lambda$-term. All of them are
described by the equation of state $\alpha_s=-1/3$. Inspite of the fact
that theoretically the string-like matter is relevant for solving the
horizon problem, it is ruled out by high redshift supernovae observations
which indicate that $\alpha_X<-0.6$ with 95 \% confidence \cite{garn98}.

Summarizing, if we want to solve the horizon problem with the aid
of the fluid with negative pressure we must assume that the $\alpha_X$
parameter is more negative than $-0.6$ and less negative than $-1$. All
values between them are so far consistent with observations. As regards
upper constraints on $\Omega_{X0}$, evidences definitely ruling out the
possibility $\Omega_{X0}>1$ (necessary for solving the horizon problem)
are not known to us. Gravitational lensing events provide such
constraints but only in the case of cosmological constant. Recent
estimate of the position of a Doppler peak in the angular power spectrum
of CMB fluctuations indicates that $0.2<\Omega_{\text{tot}}<1.5$ \cite{han98}, 
what is consistent with the presented model provided $\Omega_{m0}$ is
not too large (because $\Omega_{X0}=\Omega_{\text{tot}}-\Omega_{m0}$ must be
larger than 1). There is a hope that new instruments exploring CMB such
that VSA, MAP and Planck Surveyor satellite will provide more precise
constraints on $\Omega_{\text{tot}}$ and $\Omega_{X0}$ and will show
whether the presented scenario can be considered as a realistic physical
model. 

\acknowledgments
I thank C. van de Bruck, B. Carr, W. Kundt, A. Liddle, W. Priester and
R. Tavakol for valuable discussions.

\clearpage
\begin{figure}
\caption{Illustration of the horizon problem in the standard
cosmological model. $t_0$ is the age of the universe and $t_R$
corresponds to the recombination epoch. $r$ is a comoving radial
coordinate. Bending of photons world lines is due to the expansion.}
\label{fig1}
\end{figure}
\begin{figure}
\caption{Solution of the horizon problem in an inflationary scenario.
Due to the exponential or power law expansion of the very early universe
the world lines of photons hitting points $A$ and $B$ bend in such a
way that the light cones intersect.}
\label{fig2}
\end{figure}
\begin{figure}
\caption{Illustration of angular coordinates $\chi_R$ and $r_R$ in
a closed universe. Particle horizons of the points $A$ and $B$ at the
recombination epoch do not overlap. Dotted lines represent relic
photons paths from the last scattering surface to the observer.}
\label{fig3}
\end{figure}
\begin{figure}
\caption{Solution of the horizon problem in a closed universe.
Inspite of the fact that $\chi_R\gg r_R$ (as in a flat universe)
particle horizons of the points $A$ and $B$ at the recombination do
overlap and the horizon problem does not appear.}
\label{fig4}
\end{figure}
\begin{figure}
\caption{Admissible values of the ``abnormal'' fluid
energy density parameter $\Omega_{X0}$ solving the horizon problem
for various values of the $\alpha_X$ parameter and for different values of
the nonrelativistic matter energy density parameter $\Omega_{m0}$.}
\label{fig5}
\end{figure}

\clearpage

\begin{table}
\caption{Admissible values of $\Omega_{v0}$ for which the horizon
problem does not exist. $t_0$ is the age of the universe in units
$10^9\times h^{-1}$ years.
\label{table1}}
\begin{tabular}{cccc}
$\Omega_{m0}$ & $\Omega_{v0~min}$ & $\Omega_{v0~max}$ & $t_0$\\
\tableline
0.015 & 1.0567 & 1.0572 & 22.5 \\
0.1 & 1.165 & 1.172 & 16.1 \\
0.2 & 1.246 & 1.259 & 13.9  \\
0.3 & 1.310 & 1.327 & 12.6  \\
\end{tabular}
\end{table}

\begin{table}
\caption{ Admissible values of $\Omega_{s0}$ for which the horizon
problem does not exist. $t_0$ is the age of the universe in units
$10^9\times h^{-1}$ years.
\label{table2}}

\begin{tabular}{cccc}
$\Omega_{m0}$&$\Omega_{s0~min}$&$\Omega_{s0~max}$&$t_0$\\
\tableline
0.015 & 1.379 & 1.418 & 9.5 \\
0.1 & 1.623 & 1.685 & 8.8 \\
0.2 & 1.798 & 1.875 & 8.3  \\
0.3 & 1.933 & 2.022 & 8.0  \\
\end{tabular}
\end{table}

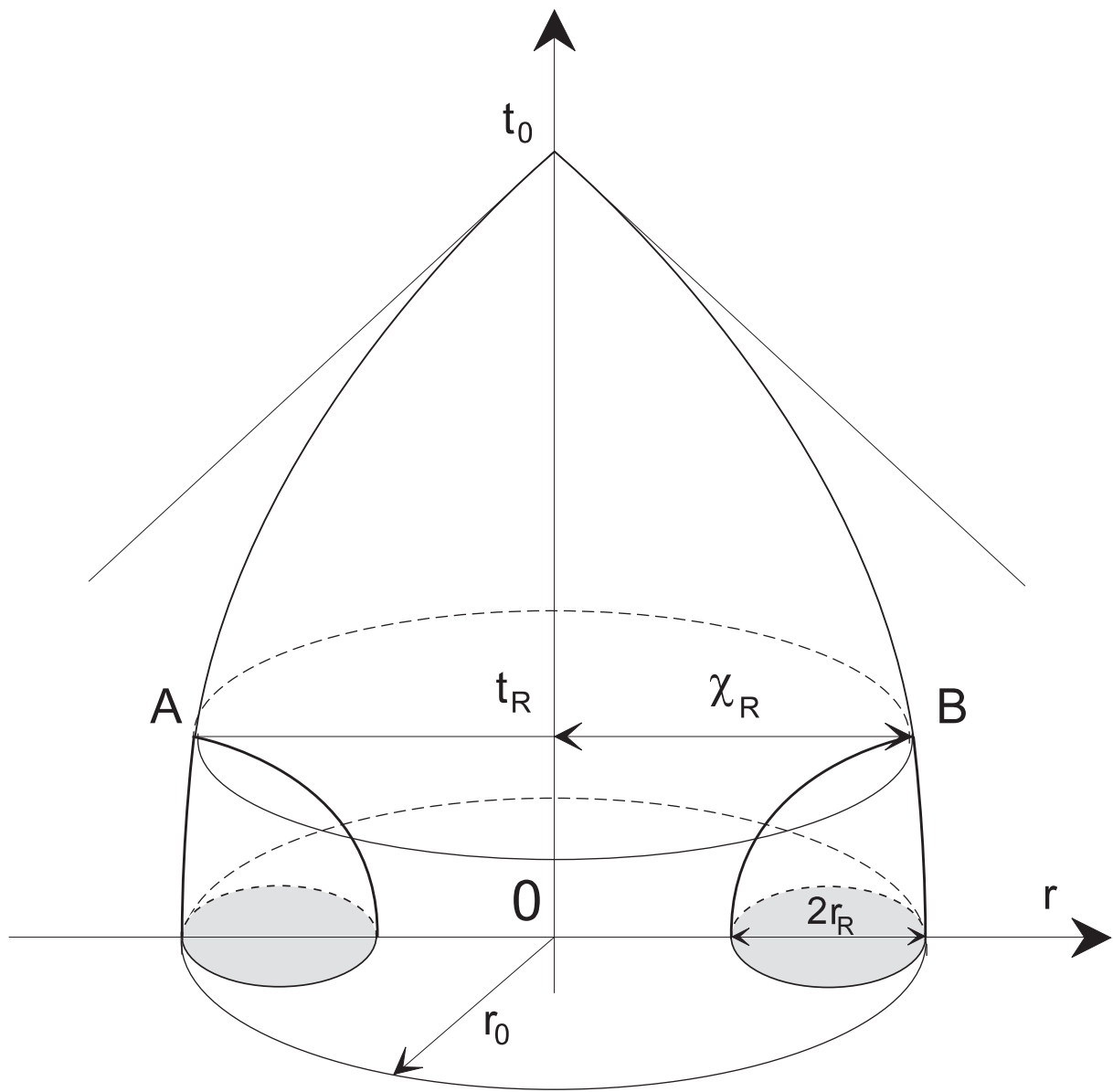
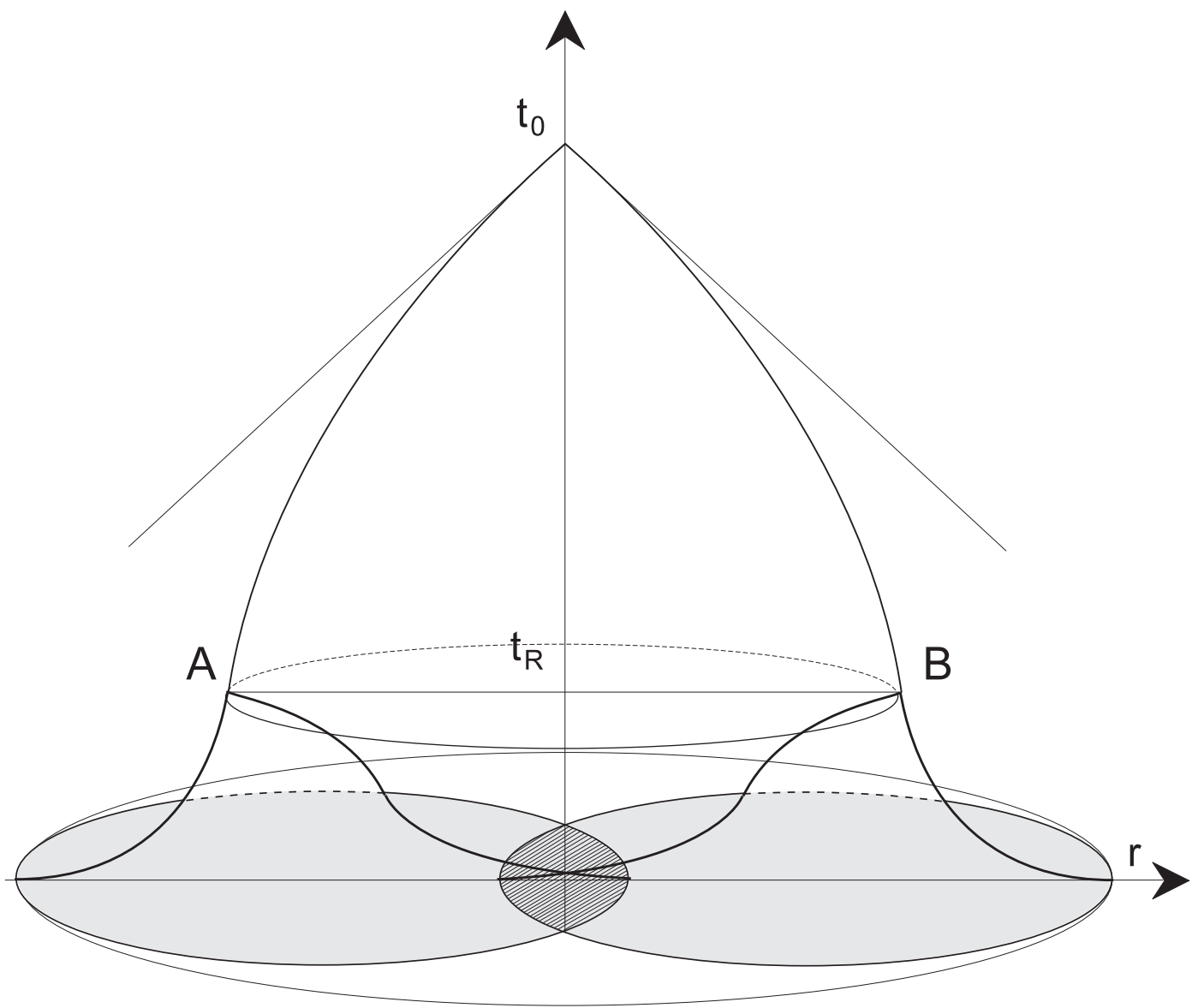
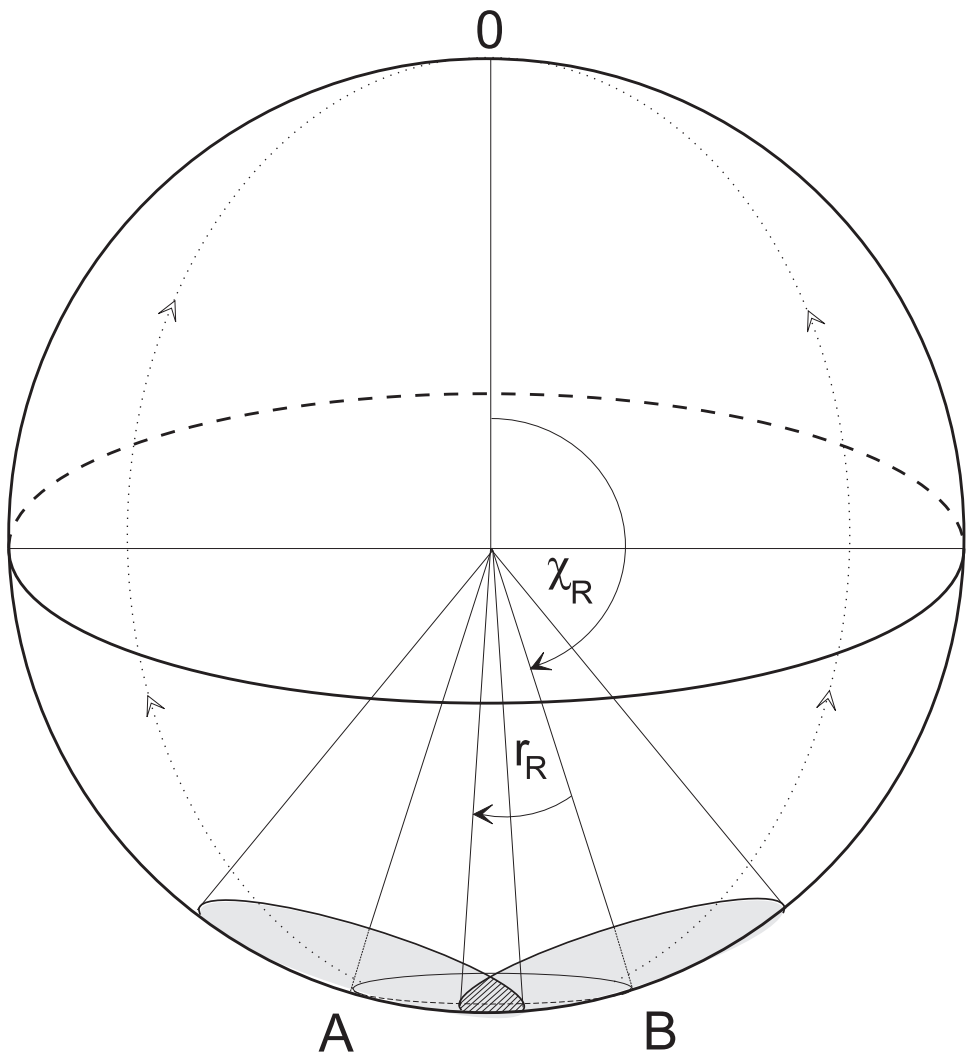
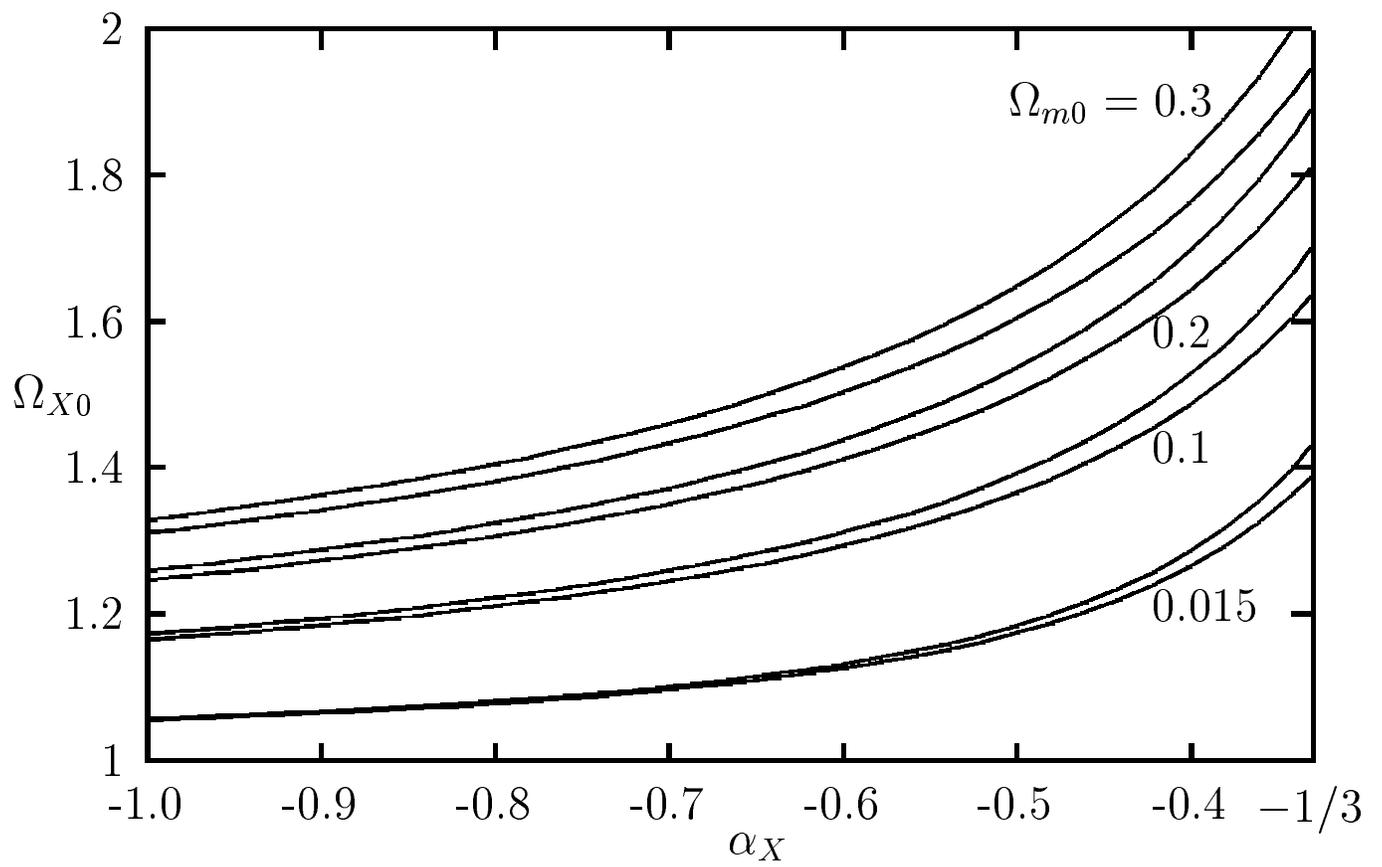
\end{document}